\documentclass[10pt,twocolumn,twoside]{IEEEtran}%
\usepackage{amsmath}
\usepackage{graphicx}
\usepackage{amsfonts}
\usepackage{amssymb}
\usepackage{epstopdf}%
\providecommand{\U}[1]{\protect\rule{.1in}{.1in}}
\providecommand{\U}[1]{\protect\rule{.1in}{.1in}}

\makeatletter

\@addtoreset{theorem}{section}

\@addtoreset{corollary}{section}

\@addtoreset{lemma}{section}

\@addtoreset{definition}{section}
\makeatother
\IEEEoverridecommandlockouts
\allowdisplaybreaks
\title{{\LARGE \textbf{Lifetime
Maximization of Wireless Sensor Networks with a Mobile Source Node}}}
\author{{\parbox{3.5in}{\center\textbf{ S. Pourazarm, C. G. Cassandras}
\\Division of Systems Engineering and\\Center for Information and Systems Engineering,\\Boston University, Boston, MA 02215\\\texttt{sepid@bu.edu, cgc@bu.edu}}}
\thanks{\footnotesize The authors’ work is supported in part by NSF under grants CNS-
1239021 and IIP-1430145, by AFOSR under grant FA9550-12-1-0113, by
ONR under grant N00014-09-1-1051, and by the Cyprus Research Promotion
Foundation under Grant New Infrastructure Project/Strategic/0308/26.}}
\begin{document}
\maketitle

\begin{abstract}
We study the problem of routing in sensor networks where the goal is to maximize the network's lifetime. Previous work has considered this problem for fixed-topology networks. Here, we add mobility to the source node, which requires a new definition of the network lifetime. In particular, we redefine lifetime to be the time until the source node depletes
its energy. When the mobile node's trajectory is unknown in advance, we formulate three versions of an optimal control problem aiming at this lifetime maximization. We show that in all cases, the solution can be reduced to a sequence of Non-Linear Programming (NLP) problems solved on line as the source node trajectory evolves and include simulation examples to illustrate our results. When the mobile node's trajectory is known in advance, we formulate an optimal control problem which, in this case, requires an explicit off-line numerical solution.
\end{abstract}

\section{Introduction}

A Wireless Sensor Network (WSN) is a spatially distributed wireless network
consisting of low-cost autonomous nodes which are mainly battery powered and
have sensing and wireless communication capabilities \cite{WSN}. Applications
range from exploration, surveillance, and target tracking, to environmental
monitoring (e.g., pollution prevention, agriculture). Power management is a
key issue in WSNs, since it directly impacts their performance and their
lifetime in the likely absence of human intervention for most applications of
interest. Since the majority of power consumption is due to the radio
component \cite{powertossim}, nodes usually rely on short-range communication
and form a multi-hop network to deliver information to a base station. Routing
schemes in WSNs aim to deliver data from the data sources (nodes with sensing
capabilities) to a data sink (typically, a base station) in an
energy-efficient and reliable way. The problem of routing in WSNs with the
goal of optimizing performance metrics that reflect the limited energy
resources of the network has been widely studied for \emph{static} (i.e.,
fixed topology) networks \cite{PERKINS01},\cite{PARKROUTING01},\cite{Xiaoyi05}%
,\cite{tassiulas},\cite{RoutingSurveyWSN}. In recent years, mobility in WSNs
has been increasingly introduced and studied \cite{Javad2012},\cite{Wang2005}
and \cite{Shah2003} with the aim of enhancing their capabilities. In fact, as
discussed in \cite{FRANCESCO}, mobility can affect different aspects of WSN
design, including connectivity, cost, reliability and energy efficiency. There
are various ways to exploit WSN mobility and incorporating it into different
network components. For instance, in \cite{Wang2005} sink mobility is
exploited and a Linear Programming (LP) formulation is proposed for maximizing
the network lifetime by finding the optimal sink node movement and sojourn
time at different nodes in the network. In \cite{Shah2003} mobile nodes
(mules) are used to deliver data to the base station. WSNs with partial mobility are studied in \cite{Srini2007}. As discussed in \cite{Raja2009}, there exist two modes for  sensor nodes mobility:  weak mobility, forced by  the death of some sensor nodes and strong mobility using an external agent \cite{Laib2005},\cite{Dantu2005}.

In this paper, we focus on the lifetime maximization problem in WSNs when
source nodes are mobile. This situation frequently arises when a mobile sensor
node is used to track one or more mobile targets or when there is a large area
to be monitored that far exceeds the range of one or more static sensors. In
the case of a fully static network the lifetime maximization problem was     
studied in \cite{Xiaoyi05} and \cite{tassiulas} by defining the WSN lifetime
as the time until the first node depletes its energy. Since it is often the
case that an optimal policy controlling a static WSN's resources leads to
individual node lifetimes being the same or almost the same as those of
others, this definition is a good characterization of the overall network's
lifetime in practice. In \cite{Xiaoyi05} routing was formulated as an optimal
control problem with controllable routing probabilities over network links and
it was shown that in a fixed network topology there exists an optimal policy
consisting of time-invariant routing probabilities. Moreover, the optimal
control problem may be converted into the LP formulation used in
\cite{tassiulas}. It is worth mentioning that a routing policy based
on probabilities can easily be implemented by transforming these probabilities
to packet flows over links and using simple mechanisms to ensure that flows
are maintained over time. In \cite{CGC2014} the simplifying assumption of idealized
batteries used as energy sources for nodes was also relaxed and a more
elaborate model was used to capture nonlinear dynamic phenomena that are known
to occur in non-ideal batteries. A somewhat surprising result was that again
an optimal policy exists which consists of time-invariant routing
probabilities and that in fact this property is independent of the parameters
of the battery model. However, this attractive property for routing is limited
to a fixed network topology.

Adding mobility to nodes raises several questions. First, one can no longer
expect that a routing policy would be time invariant. Second, it is no longer
reasonable to define the WSN lifetime in terms of the the first node depleting
its energy. For instance, if a source node travels far from some relay nodes
it was originally using, it is likely that it should no longer rely on them
for delivering data to the base station. In this scenario, the network remains
\textquotedblleft alive\textquotedblright\ even when these nodes die. Thus, in
view of node mobility, we need to revisit the definition of network lifetime.
Finally, if a routing policy is time-varying, then it has to be re-evaluated
sufficiently fast to accommodate the real-time operation of a WSN.

In the sequel, we consider mobility added to the source node and assume that any
such node travels along a trajectory that it determines and which may or may
not be known in advance. We limit ourselves to a single source
node (the case of multiple mobile source nodes depends on the exact setting and is not addressed in this paper). While
on its trajectory, the source node continuously performs sensing tasks and
generates data. Our goal is to derive an optimal routing scheme in order to
maximize the network lifetime, appropriately redefined to focus on the mobile
source node. Assuming first that the source node trajectory is not known in
advance, we formulate three optimal control problems with differences in their
terminal costs and terminal constraints and investigate how they compare in
terms of the optimal routing policy obtained, total energy consumption, and
the actual network lifetime. We will also limit ourselves to ideal battery
dynamics for all nodes. However, adopting non-ideal battery models as in
\cite{CGC2014} does not change our analysis and only complicates the
solution computation. We then consider the more challenging (from a computational perspective) problem where the
source node's trajectory is known in advance, in which case this information
can be incorporated into an optimal lifetime maximization policy.

In Section \ref{sec1}, we define the network model and the energy consumption
model is presented in Section \ref{sec2}. In Section
\ref{OptControlFormulation} we formulate the maximum lifetime optimization
problem for a WSN with a mobile source node whose trajectory is not known in
advance. Starting with a new definition for the network lifetime, we show that
the solution is a sequence of Non-Linear Programming (NLP) problems along the
source node trajectory. Numerical examples are included to
illustrate our analytical results. In Section \ref{sec5}, we consider the case
when the source node trajectory is known in advance and show that its solution
leads to a Two Point Boundary Value Problem (TPBVP).

\section{Network model}

\label{sec1}
Consider a network with $N+1$ nodes where $0$ and $N$ denote the source and
destination (base station) nodes respectively. Nodes $1,...,N-1$ act as relay
nodes to deliver data packets from the source node to the base station. We
assume the source node is mobile and travels along a trajectory with a
constant velocity while generating data packets which need to be transferred
to the fixed base through static relay nodes. First, we assume the trajectory
is not known in advance. Then, we discuss the case when the trajectory is
known in Section \ref{sec5}.
Except for the base station whose energy supply is not constrained, a limited
amount of energy is available to all other nodes. Let $r_{i}(t)$ be the
residual energy of node $i$, $i=0,\ldots,N-1$, at time $t$. The dynamics of
$r_{i}(t)$ depend on the battery model used at node $i$. Here, we assume ideal
battery dynamics in which energy is depleted linearly with respect to the
node's load, $I_{i}(t)$, i.e.,
\begin{equation}
\dot{r}_{i}(t)=-I(t) \label{batdynamic}%
\end{equation}
The distance between nodes $i$ and $j$ at time $t$ is denoted by $d_{i,j}(t)$.
Since the source node is mobile, $d_{0,j}(t)$ is time-varying for all
$j=1,...,N-1$. However, $d_{i,j}(t)=d_{i,j},\ i=1,...,N-1,\ j=2,...,N$ are
treated as time-invariant with the assumption that the source node cannot be
used as a relay, i.e., any node $i>0$ must transfer data to other relay nodes
$j>0$, $j\neq i$ or directly to the base station node $N$. The source node can
send data packets to any of the relay nodes as well as to the base station,
while relay nodes can transmit/receive data packets to/from nodes in their
transmission range. Let $O(i)$ and $I(i)$ denote the set of nodes to/from
which node $i$ can send/receive data packets respectively. Then,
$O(i)=\left\{  j:d_{i,j}\leq\tau_{i}\right\}  $ and $I(i)=\{j:d_{j,i}\leq
\tau_{j}\}$ where $\tau_{i},\ i=1,...,N-1$ denotes the transmission range of
node $i$. We define $w_{ij}(t)$ to be the routing probability of a packet
from node $i$ to node $j$ at time $t$ (equivalently, a data flow from $i$ to
$j$) and the vector $w(t)=[w_{ij}(t)]^{\prime}$ defines the control in our
problem. Let us also define $\mathbf{r}(t)=[r_{0}(t),...,r_{N-1}(t)]$ as the
vector of residual energies at time $t$. For simplicity, the data sending rate
of source node $0$ is normalized to $1$ and let $G_{i}(w)$ denote the data
packet inflow rate to node $i$. Given these definitions, we can express
$G_{i}(w)$ through the following flow conservation equations:%
\begin{align}
G_{i}(w)  &  =%
{\displaystyle\sum\limits_{k\in I(i)}}
w_{ki}(t)G_{k}(w),\text{\ }i=1,\ldots,N, \  G_{0}(w) =1\label{HH01}
\end{align}
\section{Energy Consumption Model}
\label{sec2} In our WSN environment, the battery workload $I(t)$ is due to
three factors: the energy needed to sense a bit, $E_{sense}$, the energy
needed to receive a bit, $E_{rx}$, and the energy needed to transmit a bit,
$E_{tx}$. If the distance between two nodes is $d$, we have: 
$E_{tx}=p(d),\text{ \ }E_{rx}=C_{r},\text{ \ }E_{sense}=C_{e}$,  
where $C_{r}$, $C_{e}$ are given constants dependent on the communication and
sensing characteristics of nodes, and $p(d)\geq0$ is a function monotonically
increasing in $d$; the most common such function is $p(d)=C_{f}+C_{s}d^{\beta
}$ where $C_{f}$, $C_{s}$ are given constants and $\beta$ is a constant
dependent on the medium involved. We will use the common situation where
$\beta=2$ in the rest of the paper, but this has no effect on our approach. We
shall use this \ energy model, but ignore the sensing energy $C_{e}$, i.e.,
set $C_{e}=0$ (otherwise, $C_{e}$ is simply added to the source node's
workload without affecting the analysis). Clearly, this is a relatively simple
energy model that does not take into consideration the channel quality or the
Shannon capacity of each wireless channel. The ensuing optimal control
analysis is not critically dependent on the exact form of the energy
consumption model attributed to communication, although the ultimate optimal
value of $w(t)$ obviously is. 
For any node $i=1,\ldots,N-1$, the workload $I_{i}(t)$ at that node is given
by%
\begin{equation}
I_{i}(w(t))=G_{i}(w(t))[  \sum_{j\in O(i)}w_{ij}(t)(C_{s}d_{i,j}^{\beta
}(t)+C_{f})+C_{r}]  \label{I1}%
\end{equation}
and the workload $I_{0}(t)$ at the source node $0$ (recalling that
$G_{0}(w(t))=1$) is given by%
\begin{equation}
I_{0}(w(t))=\sum_{j\in O(0)}w_{0j}(t)(C_{s}d_{0,j}^{\beta}(t)+C_{f})
\label{I0}%
\end{equation}
Assuming an ideal battery behavior for all nodes as in (\ref{batdynamic}), the
state variables for our problem are $r_{i}(t),\ i=0,...,N-1$. Note that
$d_{0,j}(t)=\Vert(x_{0}(t),y_{0}(t))-(x_{j},y_{j})\Vert$, the Euclidean
distance of the source node from any other node is known at any time instant
$t$ (but not in advance) as determined by the source node's trajectory.
Finally, observe that by controlling the routing probabilities $w_{ij}(t)$ in
(\ref{I0}) and (\ref{I1}) we directly control node $i$'s battery discharge process.

\section{Optimal control problem formulation\label{OptControlFormulation}}

Our objective is to maximize the WSN lifetime by controlling the routing
probabilities $w_{ij}(t)$. For a static network, where all nodes including the
source node are fixed, the network lifetime is usually defined as the time
until the first node depletes its battery \cite{CGC2014}, \cite{tassiulas}.
However, when the source node is mobile, this definition of network lifetime
is no longer appropriate as explained in Section I. In the sequel, we
formulate three optimal control problems for maximizing lifetime in a WSN with a mobile source node and investigate their relative effect in
terms of an optimal routing policy, total energy consumption, and the network lifetime.

\subsection{Optimal Control Problem - I}

\label{f1}

We define the network lifetime as the time when the
source node runs out of energy. Consider a fixed time $t$ when the source node
is at position $(x_{0}(t),y_{0}(t))\in\mathbb{R}^{2}$. In the absence of any
future information regarding the position of this node (e.g., the node may
actually stop for some time interval before moving again), the routing problem
we face is one of a fixed topology WSN similar to the one in \cite{CGC2014} but
with different terminal state constraints due to the new network lifetime
definition. Thus, this maximum lifetime optimal control problem is formulated
as follows, using the variables defined in (\ref{HH01}), (\ref{I1}) and (\ref{I0}):%
\begin{gather}
\min_{w(t)}-\int_{0}^{T}dt\label{obj}\\
\text{s.t.\ \ }\dot{r}_{i}(t)=-I_{i}(w(t)),\   r_{i}(0)=R_{i},\quad
i=0,..,N-1\label{riC4}\\
I_{i}(w(t))=G_{i}(w(t))[\sum_{j\in O(i)}w_{ij}(t)(C_{s}d_{i,j}^{2}%
+C_{f})+C_{r}],\nonumber\\
\quad\quad i=1,...,N-1\label{Ii}\\
I_{0}(w(t))=\sum_{j\in O(0)}w_{0j}(t)(C_{s}d_{0,j}^{2}(t)+C_{f})\label{I_0}\\
d_{0,j}(t)=\Vert(x_{0}(t),y_{0}(t))-(x_{j},y_{j})\Vert,\ x_{0}(t),y_{0}%
(t)\text{ given}\nonumber\\
G_{i}(w(t))=%
{\displaystyle\sum\limits_{k\in I(i)}}
w_{ki}(t)G_{k}(w(t)),\text{ }i=1,..,N-1\label{Gi}\\
\sum_{j\in O_{i}}w_{ij}(t)=1,\quad0\leq w_{ij}(t)\leq1,\text{ }i=0,\ldots
,N-1\label{wij}\\
r_{0}(T)=0\label{riatT}\\
r_{0}(t)>0,\ t\in\lbrack0,T); \text{ }r_{i}(t)\geq0,\text{ }i=1,..,N-1,\text{ }t\in\lbrack0,T]
\label{r_i}%
\end{gather}
where $r_{i}(t)$, $i=0,...,N-1$, are the state variables representing the node
$i$ battery dynamics and $(x_{0}(t),y_{0}(t))$ are the given instantaneous
coordinates of the source node at time $t$. Control constraints are specified
through (\ref{wij}). Finally, (\ref{riatT}) provides the boundary conditions
for $r_{0}(t)$ at $t=T$ requiring that the terminal time is the time when the
source node depletes its energy. 
 In what follows,
we will use $w^{\ast}(t)$ to denote the optimal routing vector at the fixed
time $t$.

\subsubsection{Optimal control problem I solution\label{OptControlSolution}}

We begin with the Hamiltonian for this optimal control problem:
\begin{align}
&H(\mathbf{r},x_{0},y_{0},w,\lambda,t)=&-1+\lambda_{0}(t)(-I_{0}(t))+\nonumber\\&&\sum
_{i=1}^{N-1}\lambda_{i}(t)(-I_{i}(t)) \label{Ham1}%
\end{align}
where $\lambda_{i}(t)$ is the costate corresponding to $r_{i}%
(t),\ i=0,...,N-1$ and must satisfy:
\begin{equation}
\dot{\lambda}_{i}(t)=-\dfrac{\partial H}{\partial r_{i}}=0\quad i=0,...,N-1
\label{costates}%
\end{equation}
Therefore, $\lambda_{i}$, $i=0,...,N-1$, are constants. To determine their
values we make use of the boundary conditions which follow from (\ref{riatT}),
i.e., the terminal state constraint function is $\Phi(\mathbf{r}(T))=\nu
r_{0}(T)$ and the costate boundary conditions are given by:%
\[
\lambda_{i}(T)=\dfrac{\partial\Phi(r_{0}(T),...,r_{N-1}(T))}{\partial
r_{i}(T)},\quad i=0,...,N-1
\]
which implies that
\begin{align}
\lambda_{i}    =0\quad i=1,...,N-1,\quad
\lambda_{0}    =\nu\label{lambdaT1}%
\end{align}
where $\nu$ is some scalar constant. Finally, the optimal solution must
satisfy the transversality condition $H(T)+\left.  {\partial\Phi
}/{\partial t}\right\vert _{t=T}=0$, i.e., $
-1+\nu\dot{r}_{0}(T)+\nu\dot{r}_{0}(T)=0$, 
which yields:
$\nu={1}/{2\dot{r}_{0}(T)}<0 $,  
where the inequality follows from (\ref{riatT}) and (\ref{r_i}) which imply
that $\dot{r}_{0}(T)<0$ and consequently $\nu<0$.

\textbf{Theorem 1:} There exists a time-invariant solution of (\ref{obj}%
)-(\ref{r_i}): $w^{\ast}(t)=w^{\ast}(T)$, $t\in\lbrack0,T]$.

\emph{Proof:} Observe that the control variables $w_{ij}(t)$ appear in the
problem formulation (\ref{obj})-(\ref{r_i}) only through $I_{i}(w(t))$.
Applying the Pontryagin minimum principle to (\ref{Ham1}):
\[
\lbrack I_{0}^{\ast}(t),...,I_{N-1}^{\ast}(t)]=\arg{\min_{I_{i}\geq
0;\ i=0,...,N-1}H(I_{i},t,\lambda^{\ast})}%
\]
and making use of the fact that we found $\lambda_{i}=0,\ i=1,...,N-1$, we
have: 
$
I_{0}^{\ast}(t)=\arg{\min_{I_{0}(t)>0}(-1-\nu I_{0}(t)}) $. 
Recalling that $\nu<0$, in order to minimize the  Hamiltonian, we
need to minimize $I_{0}(t)$. Therefore, the optimal control problem
(\ref{obj})-(\ref{r_i}) is reduced to the following optimization problem which
we refer to as $\mathbf{P1(}t\mathbf{)}$:
\begin{gather}
\min_{w(t)}I_{0}(t)\label{obj2}\\
\text{s.t.   } I_{i}(w(t))=G_{i}(w(t))[\sum_{j\in O(i)}w_{ij}(t)(C_{s}d_{i,j}^{2}%
+C_{f})+C_{r}],\nonumber\\
\quad\quad i=1,...,N-1\label{Ii2}\\
I_{0}(w(t))=\sum_{j\in O(0)}w_{ij}(t)(C_{s}d_{0,j}^{2}(t)^{2}+C_{f}%
)\label{I02}\\
d_{0,j}(t)=\Vert(x_{0}(t),y_{0}(t))-(x_{j},y_{j})\Vert,\ \ x_{0}%
(t),y_{0}(t)\text{ given}\nonumber\\
G_{i}(w(t))=%
{\displaystyle\sum\limits_{h\in I(i)}}
w_{hi}(t)G_{h}(w),\text{ }i=1,\ldots,N-1\label{Gi2}\\
\sum_{j\in O(i)}w_{ij}(t)=1,\quad0\leq w_{ij}(t)\leq1,i=0,..
,N-1\label{wij2}\\
\int_{0}^{T}I_{0}(t)dt=R_{0} \label{TerminalCondition}%
\end{gather}
When $t=T$, the solution of this problem is $w^{\ast}(T)$ and depends only on
the fixed network topology and the values of the fixed energy parameters in
(\ref{I02}) and the control variable constraints (\ref{wij2}). The same
applies to any other $t\in\lbrack0,T)$, therefore, there exists a
time-invariant optimal control policy $w^{\ast}(t)=w^{\ast}(T)$, which
minimizes the Hamiltonian and proves the theorem. $\blacksquare$

As the trajectory of the source node evolves, let us discretize it using a
constant time step $\delta$. Thus, at time instants $t,t+\delta,\ldots
,t+k\delta,\ldots$ we solve problems $\mathbf{P1(}t+k\delta\mathbf{)}$,
$k=0,1,\ldots$ based on the associated source node positions $(x_{0}%
(t+k\delta),y_{0}(t+k\delta))$, $k=0,1,\ldots$ as they become available.
Theorem 1 asserts that at each time step, there exists a fixed optimal routing
vector $w^{\ast}(t+k\delta)\equiv w_{k}^{\ast}$ associated with the source
node's position. Thus, an optimal routing vector at each time step is obtained
by solving the following NLP:
\begin{gather}
\min_{w^{k}}\sum_{j\in O^{k}(0)}w_{0j}^{k}(C_{s}(d_{0,j}^{k})^{2}%
+C_{f})\label{obj3}\\
\text{s.t.  }\sum_{j\in O^{k}(i)}w_{ij}^{k}=1,\quad0\leq w_{ij}^{k}\leq1,i=0,..,N-1
\label{wij3}%
\end{gather}
where $w^{k}$ is a routing vector at step $k$, $O^{k}(i)$ is the set of output
nodes of $i$ (which may have changed since some relay nodes may have died),
and $d_{0,j}^{k}=\Vert(x_{0}^{k},y_{0}^{k})-(x_{j},y_{j})\Vert$ is the
distance between the source node and node $j$ at the $k$th step. Observe that
in (\ref{obj3}) the objective value is minimized over $w^k_{0j}$, $j\in
O^{k}(0)$ leaving the remaining routing probabilities $w^k_{ij}%
,\ i=1,...,N-1,\ j\in O^k(i)$, subject only to the feasibility constraints
(\ref{wij3}). Therefore, at each iteration, the source node sends data packets
to its nearest neighbors in $O^k(0)$ in order to minimize its load. The
remaining routing probabilities need to be feasible according to (\ref{wij3}).
The simplest such feasible solution is obtained by sending the inflow data
packets to the neighbors of a relay node uniformly, i.e., $w^k_{ij}=\dfrac
{1}{|O^k(i)|},\ i=1,...,N-1$. Finally, at the end of each iteration we update
the residual energy of all nodes as follows:
\begin{equation}
r_{i}^{k+1}=r_{i}^{k}-I_{i}(w^{k})\cdot\delta\label{rical}%
\end{equation}
If $r_{0}^{k+1}\leq0$ we declare the network to be dead. However, if
$r_{i}^{k+1}\leq0,\ i=1,..,N-1$, then we omit dead nodes and update the
network topology to calculate $w_{k+1}^{\ast}$ in the next iteration with one
fewer node. Note that it is possible for all relay nodes to be dead while
$r_{0}^{k+1}>0$, implying that the source node still has the opportunity to
transmit data directly to the base if $N\in O^{k+1}(0)$.

The fact that the solution of $\mathbf{P1(}t\mathbf{)}$ does not allow any
direct control over the relay nodes is a drawback of this formulation and
motivates the next definition of WSN lifetime.

\subsection{Optimal Control Problem - II}

As already mentioned, the optimization problem (\ref{obj3})-(\ref{wij3}) does
not directly control the way relay nodes consume their energy. To impose such
control on their energy consumption, we add $\sum_{i=1}^{N-1}r_{i}(T)$ as a
terminal cost to the objective function of the optimal control problem
(\ref{obj})-(\ref{r_i}) and formulate a new problem as follows:
\begin{gather}
 \min_{w(t)}\left(  -\int_{0}^{T}dt+\epsilon\sum_{i=1}^{N-1}r_{i}(T)\right)\ \ 
\text{s.t.}\quad(\ref{riC4})-(\ref{r_i})\label{obj4} 
\end{gather}
where $\epsilon>0$ is a weight reflecting the importance of the total residual
energy relative to the lifetime as measured at time $t$. Thus, in order to
minimize the terminal cost, relay nodes are compelled to drive their residual
energy to be as close to zero as possible at $t=T$. This plays a role as we
solve the sequence of problems resulting for the source node movement: the
inclusion of this terminal cost tends to preserve some relay node energy which
may become important in subsequent time steps. The solution of (\ref{obj4})
obviously results in a different network lifetime $T^{\ast}$ relative to that
of problem (\ref{obj})-(\ref{r_i}), which is recovered when $\epsilon=0$.

\subsubsection{Optimal control problem II solution}

The Hamiltonian based on the new objective function (\ref{obj4}), as well as
the costate equations, are the same as (\ref{Ham1}) and (\ref{costates})
respectively. However, the terminal state constraint is now
\[
\Phi(\mathbf{r}(T))=\epsilon\sum_{i=1}^{N-1}r_{i}(T)+\nu r_{0}(T)
\]
and the costate boundary conditions are given by:%
\[
\lambda_{i}(T)=\dfrac{\partial\Phi(r_{0}(T),...,r_{N-1}(T))}{\partial
r_{i}(T)},\quad i=0,...,N-1
\]
so that $
\lambda_{i}=\epsilon,\quad i=1,...,N-1 \quad  \lambda_{0}=\nu$. 
Finally, the transversality condition $H(T)+\left.  {\partial\Phi
}/{\partial t}\right\vert _{t=T}=0$ for this problem is%
\[
-1+\nu\dot{r}_{0}(T)+\epsilon \sum_{i=1}^{N-1}\dot{r}_{i}(T)+\nu\dot{r}_{0}%
(T)+\epsilon\sum_{i=1}^{N-1}\dot{r}_{i}(T)=0
\]
resulting in
\begin{equation}
\nu=\dfrac{1-2 \epsilon\sum_{i=1}^{N-1}\dot{r}_{i}(T)}{2\dot{r}_{0}(T)}\leq0
\label{nu1}%
\end{equation}
Looking at (\ref{riatT}) and (\ref{r_i}) and as already discussed in the
previous section, we have $\dot{r}_{0}(T)<0$. For the any relay node
$i=1,...,N-1$, there are two possible cases: $(i)$ Node $i$ is not
transmitting any data at $t=T$, i.e., the node is already out of energy or the
inflow rate to that node is zero, $G_{i}(w(T))=0$. In this case, $I_{i}(T)=0$,
consequently $\dot{r}_{i}(T)=0$. $(ii)$ Node $i$ is transmitting, i.e.,
$I_{i}(T)>0$, therefore, $\dot{r}_{i}(T)<0$. It follows that $\sum_{i=0}%
^{N-1}\dot{r}_{i}(T)\leq0$ and we conclude that $\nu\leq0$.

\textbf{Theorem 2:} There exists a time-invariant solution of (\ref{obj4}):
$w^{\ast}(t)=w^{\ast}(T)$, $t\in\lbrack0,T]$.

\emph{Proof: } The proof is similar to that of Theorem 1. First, observe that
the control variables $w_{ij}(t)$ appear in the problem formulation
(\ref{obj4}) only through $I_{i}(w(t))$. Next, applying the Pontryagin minimum
principle to (\ref{Ham1}) and based on our analysis we get:
\begin{equation}
\lbrack I_{0}^{\ast}(t),...,I_{N-1}^{\ast}(t)]=\arg\min_{I_{i}(t)\geq0}[-1-\nu
I_{0}(t)-\epsilon\sum_{i=1}^{N-1}I_{i}(t)] \label{Pont2}%
\end{equation}
Recalling that $\nu\leq0$ in (\ref{nu1}), in order to minimize (\ref{Pont2})
the routing vector should minimize $I_{0}(t)$ while maximizing $\epsilon
\sum_{i=1}^{N-1}I_{i}(t)$. Therefore, the optimal control problem (\ref{obj4})
can be written as the following problem $\mathbf{P2(}t\mathbf{)}$:
\begin{gather}
\min_{w(t),\nu}(  I_{0}(t)+\frac{\epsilon}{\nu}\sum_{i=1}^{N-1}%
I_{i}(t)) \ \ 
\text{s.t.}\quad(\ref{Ii2})-(\ref{TerminalCondition})\label{obj6}
\end{gather}
where $\nu<0$ is an unknown constant which must also be determined (if $\nu
=0$, the problem in (\ref{Pont2}) reduces to maximizing $\epsilon\sum
_{i=1}^{N-1}I_{i}(t)$ and can be separately solved). Using the same argument
as in Theorem 1, at $t=T$, the solution $w^{\ast}(T)$ depends only on the
fixed network topology and the values of the fixed energy parameters in
(\ref{I02}) and the control variable constraints (\ref{Ii2})-(\ref{wij2}). The
same applies to any other $t\in\lbrack0,T)$, therefore, there exists a
time-invariant optimal control policy $w^{\ast}(t)=w^{\ast}(T)$, which
minimizes the Hamiltonian and proves the theorem. $\blacksquare$

The intuition behind $\mathbf{P2(}t\mathbf{)}$ in (\ref{obj6}) is that one can
prolong the network lifetime by minimizing the load of the source node while
maximizing the workload of relay nodes. As in the case of $\mathbf{P1(}%
t\mathbf{)}$, we proceed by discretizing the source node trajectory and
determining at step $k$ an optimal routing vector $w_{k}^{\ast}$ and
associated $\nu_{k}^{\ast}$ by solving the NLP:
\begin{gather}
\min_{w^{k},\nu^{k}}\left(  I_{0}(w^{k})+\frac{\epsilon}{\nu^k}\sum_{i=1}%
^{N-1}I_{i}(w^{k})\right) \ \ \text{s.t. (\ref{nu1}) and}\label{obj7}\\
I_{i}(w^{k})=G_{i}(w^{k})[\sum_{j\in O^{k}(i)}w_{ij}^{k}(C_{s}d_{i,j}%
^{2}+C_{f})+C_{r}]\label{Ii7}\\
I_{0}(w^{k})=\sum_{j\in O^{k}(0)}w_{ij}^{k}(C_{s}(d_{0,j}^{k})^{2}%
+C_{f})\label{I07}\\
G_{i}(w^{k})=%
{\displaystyle\sum\limits_{h\in I^{k}(i)}}
w_{hi}^{k}G_{h}(w^{k}),\text{ }i=1,\ldots,N-1\label{Gi7}\\
\sum_{j\in O^{k}(i)}w_{ij}^{k}=1,\quad0\leq w_{ij}^{k}\leq1,\text{ }%
i=0,\ldots,N-1 \label{wij7}%
\end{gather}
We then evaluate the energy level of all nodes using (\ref{rical}) and check
the terminal constraint (\ref{riatT}) at the end of each iteration. If the
source node is \textquotedblleft alive\textquotedblright, we update the
network topology to eliminate any relay nodes that may have depleted their
energy in the current time step. Note that in order to solve (\ref{obj7}%
)-(\ref{wij7}) we also need to determine $\nu^{k}$ so that it satisfies
(\ref{nu1}) with $\dot{r}_{i}^{\ast}(T)=-I_{i}(w^{\ast}(T))=-I_{i}(w_{k}%
^{\ast})$. To do so, we start with an initial value and iteratively update it
until (\ref{nu1}) is satisfied. This extra step adds to the problem's
computational complexity and motivates yet another definition of WSN lifetime.

\subsection{Optimal Control Problem - III}

In this section, we revise the terminal constraint used in Problem I in order
to improve the total energy consumption in the network and possibly reduce the
computational effort required in $\mathbf{P2(}t\mathbf{)}$ due to the presence
of $\nu$ in (\ref{obj6}). Thus, let us replace the terminal constraint
(\ref{riatT}), i.e., $r_{0}(T)=0$, by $\sum_{i=0}^{N-1}r_{i}(T)=0$, thus
redefining the WSN lifetime as the time when \emph{all} nodes deplete their energy.
Compared to Problem II where we included $\sum_{i=1}^{N-1}r_{i}(T)$ as a
\emph{soft} constraint on the total residual relay node energy, here we impose
it as a \emph{hard} constraint. The following result asserts that the source
node 0 must still die at $t=T$, just as in Problem I.

\textbf{Lemma 1:} Consider (\ref{obj})-(\ref{r_i}) with (\ref{riatT}%
)-(\ref{r_i}) replaced by $\sum_{i=0}^{N-1}r_{i}(T)=0$. Then, $\dot{r}%
_{0}(T)<0$.

\emph{Proof: }Proceeding by contradiction, suppose $\dot{r}_{0}(T)=0$,
consequently $r_{0}(t_{1})=0$ for some $t_{1}<T$ and there must exist some
node $i>0$ such that $r_{i}(t_{1})>0$ otherwise the network would be dead at
$t_{1}<T$. Then, $w_{0j}(t_{1})=0$. This implies that $G_{j}(w(t_{1}))=0$ for
all $j\in O(0)$, i.e., there is no inflow to process at any node $j\in O(0)$,
therefore, $G_{i}(w(t_{1}))=0$ at all nodes $i>0$ contradicting the fact that
$r_{i}(t_{1})>0$ for some $i>0$. $\blacksquare$

\subsubsection{Optimal control problem III solution}

We apply the new terminal constraint to problem (\ref{obj})-(\ref{r_i}),
i.e., replace (\ref{riatT})-(\ref{r_i}) by%
\begin{equation}
\sum_{i=0}^{N-1}r_{i}(T)=0 \label{TermConstr3}%
\end{equation}
The Hamiltonian is still the same as (\ref{Ham1}) and the costate equations
remain as in (\ref{costates}). However, the terminal state constraint, as well
as the costate boundary conditions, are modified as follows:
\begin{align}
&  \Phi(\mathbf{r}(T))=\nu\sum_{i=0}^{N-1}r_{i}(T)\label{phi3}\\
&  \lambda_{i}(T)=\nu\dfrac{\partial\Phi(r_{0}(T),...,r_{N-1}(T))}{\partial
r_{i}(T)}=\nu,\  i=0,...,N-1 \label{lambdaT_3}%
\end{align}
Thus, the costates over all $t\in\lbrack0,T]$ are identical constants,
$\lambda_{0}(t)=...=\lambda_{N-1}(t)=\nu$. Similar to our previous analysis,
we use the transversality condition $H(T)+\left.  {\partial\Phi
}/{\partial t}\right\vert _{t=T}=0$ to investigate the sign of $\nu$: $
-1+\sum_{i=0}^{N-1}\nu\dot{r}_{i}(T)+\nu\sum_{i=0}^{N-1}\dot{r}_{i}(T)=0$  and 
 we get $
\nu=\dfrac{2}{\sum_{i=0}^{N-1}\dot{r}_{i}(T)}\leq0$  
by examining all possible cases for the state of relay nodes at $t=T$ as we
did for (\ref{nu1}). Finally, applying the Pontryagin minimum principle leads
to the following optimization problem $\mathbf{P3(}t\mathbf{)}$:
\begin{gather}
\min_{w(t)}\sum_{i=0}^{N-1}I_{i}(t)\ \ 
\text{s.t.}\quad(\ref{Ii2})-(\ref{wij2}) \text{ and }\  \label{obj8}\\
\sum_{i=0}^{N-1}\int_{0}^{T}I_{i}(t)dt=\sum_{i=0}^{N-1}R_{i}
\label{TerminalCondition2}%
\end{gather}
This new formulation indicates that the optimal routing vector corresponds to
a policy minimizing the overall network workload during its lifetime, $T$. We
can once again establish the fact that there exists a time-invariant solution
of (\ref{obj8})-(\ref{TerminalCondition2}) $w^{\ast}(t)=w^{\ast}(T)$,
$t\in\lbrack0,T]$ with similar arguments as in Theorems 1 and 2, so we omit
this proof. We then proceed as before by discretizing the source node
trajectory and determining at step $k$ an optimal routing vector $w_{k}^{\ast
}$ by solving the NLP:
\begin{gather}
\min_{w^{k}}\sum_{i=0}^{N-1}I_{i}(w^{k})\ \ 
\text{s.t.}\quad(\ref{Ii7})-(\ref{wij7})\label{obj9}
\end{gather}
Note that problem (\ref{obj9}\textbf{) }is not always feasible. In fact, its
feasibility depends on the initial energies of the nodes $r_{i}(0)=R_{i},$
$i=0,..,N-1,$ in (\ref{riC4}). It was shown in \cite{CGC2014} that, for a
fixed network topology, if we can optimally allocate initial energies to all
nodes, this results in all nodes dying simultaneously, which is exactly what
(\ref{TermConstr3}) requires. However, such degree of freedom does not exist
in (\ref{obj9}\textbf{)}, therefore, one or more instances of (\ref{obj9}%
\textbf{)} for $k=0,1,\ldots$ is likely to lead to an infeasible NLP problem
since we cannot control $R_{i}^{k}$. Clearly, this makes the definition of
WSN\ lifetime through (\ref{TermConstr3}) undesirable. Nonetheless, we follow
up on it for the following reason: We will show next that (\ref{obj9}), if
feasible, is equivalent to a shortest path problem and this makes it extremely
efficient for on-line solution at each time step along the source node
trajectory. Thus, if we adopt a shortest path routing policy at every step $k$,
even though it is no longer guaranteed that this solves (\ref{obj9}) since
(\ref{TermConstr3}) may not be satisfied for the values of $R_{i}^{k}$ at this
step, we can still update all node residual energies through (\ref{riC4}) and
check whether $r_{0}^{k+1}\leq0$. The network is declared dead as soon as this
condition is satisfied, even if $\sum_{i=0}^{N-1}r_{i}^{k+1}>0$. Although
(\ref{TermConstr3}) is not satisfied at the $k$th step, this approach provides
a computationally efficient heuristic for maximizing the WSN lifetime over the
source node trajectory in the sense that when $r_{0}^{k+1}\leq0$ at time
$t+k\delta$, the lifetime is $T=t+k\delta$ and this may compare favorably to
the solution obtained through the Problem II formulation where both lifetimes
satisfy $r_{0}(T)=0$ with $\dot{r}_{0}(T)<0$ (by Lemma 1). This idea is tested
in Section \ref{NumExamples}.

\subsubsection{Transformation of Problem III to a shortest path problem}

 The WSN can be
modeled as a directed graph from the source (node 0) to a destination (node
$N$). Each arc $(i,j)$ is a transmission link from node $i$ to node $j$. The
weight of arc $(i,j)$ is defined as 
$
Q_{ij}=C_{r}+C_{s}\cdot d_{i,j}^{2}+C_{f}%
$  
which is the energy consumption to transmit one bit of information from node
$i$ to node $j$. A path from the source to the destination node is denoted by
$p$ with an associated cost defined as 
$
C_{p}=\sum_{(i,j)\in p}Q_{ij}$. 
Clearly, for each bit of information, the total energy cost to deliver it from
the source node to the base station through path $p$ is $C_{p}$.

\textbf{ Theorem 3:} If problem (\ref{obj9}) is feasible, then its solution is
equivalent to the shortest path on the graph weighted by the transmission
energy costs $Q_{ij}$ for each arc $(i,j)$.

\emph{Proof: }We first prove that if the solution of (\ref{obj9}) includes
multiple paths from node 0 to $N$ where nodes in the path have positive
residual energy, then the paths have the same cost. We proceed using a
contradiction argument. Suppose that in the optimal solution there exist two
distinct paths $P_{1}^{\ast}$ and $P_{2}^{\ast}$ such that $C_{P_{1}^{\ast}%
}<C_{P_{2}^{\ast}}$. Let $q_{P_{1}^{\ast}}$ and $q_{P_{2}^{\ast}}$ be the
amounts of information transmitted through $P_{1}$ and $P_{2}$ respectively in
a time step of length $\delta$, i.e., $q_{P_{1}^{\ast}}+q_{P_{2}^{\ast}}%
=G_{0}\cdot\delta$.

In addition, let $\bar{r}_{k}^{\ast}$ be the total amount of energy consumed
under an optimal routing vector $w_{k}^{\ast}$ over the time step of length
$\delta$, i.e., $\bar{r}_{k}^{\ast}=\sum_{i}I_{i}(w_{k}^{\ast})\cdot\delta$.
It follows that
$
q_{P_{1}^{\ast}}C_{P_{1}^{\ast}}+q_{P_{2}^{\ast}}C_{P_{2}^{\ast}}=\bar{r}%
_{k}^{\ast}%
$. 
Suppose we perturb the optimal solution so that an additional amount of
data $\epsilon>0$ is transmitted through $P_{1}^{\ast}$. Then:
\[
(q_{P_{1}^{\ast}}+\epsilon)C_{P_{1}^{\ast}}+(q_{P_{2}^{\ast}}-\epsilon
)C_{P_{2}^{\ast}}=\bar{r}_{k}^{\ast}+\epsilon(C_{P_{1}^{\ast}}-C_{P_{2}^{\ast
}})<\bar{r}_{k}^{\ast}%
\]
This implies that $\sum_{i}I_{i}(w_{k}^{\ast})$ is not the minimum cost and
the original solution is not optimal, leading to a contradiction.

We have thus established  that if the solution of (\ref{obj9}) (if it
exists) includes multiple paths from node 0 to $N$ where nodes in the path
have positive residual energy, then the paths have the same cost. Recall that
arc weights correspond to energy consumed, therefore the shortest path on the
graph weighted by the transmission energy costs guarantees the lowest cost to
deliver every bit of data from the source node to the base station, i.e.,
$\min\sum_{i=0}^{N-1}I_{i}(w^{k})$. $\blacksquare$

\subsection{Numerical examples\label{NumExamples}}

\begin{figure}[ptbh]
\centering
\includegraphics[scale=0.48]{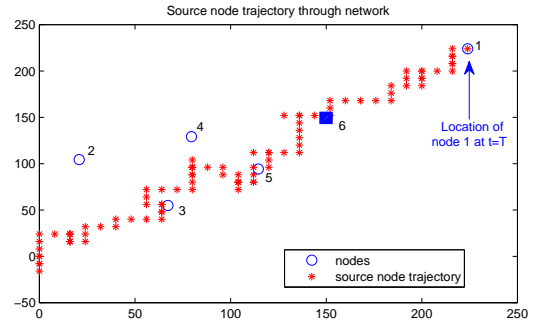}\caption{Example 6-node network with
mobile source (node 1)}%
\label{TPLG1}%
\end{figure}

\begin{figure}[ptbh]
\centering
\includegraphics[scale=0.57]{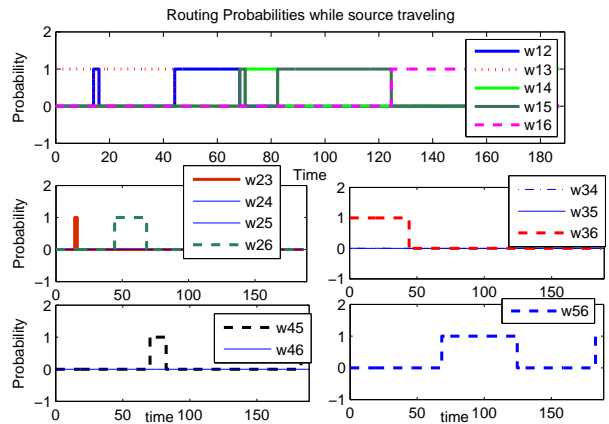}\caption{Optimal routing vector for the WSN
in Fig. \ref{TPLG1} with a mobile source node (Problem II)}%
\label{RV}%
\end{figure}

\begin{figure}[ptbh]
\centering
\includegraphics[scale=0.38]{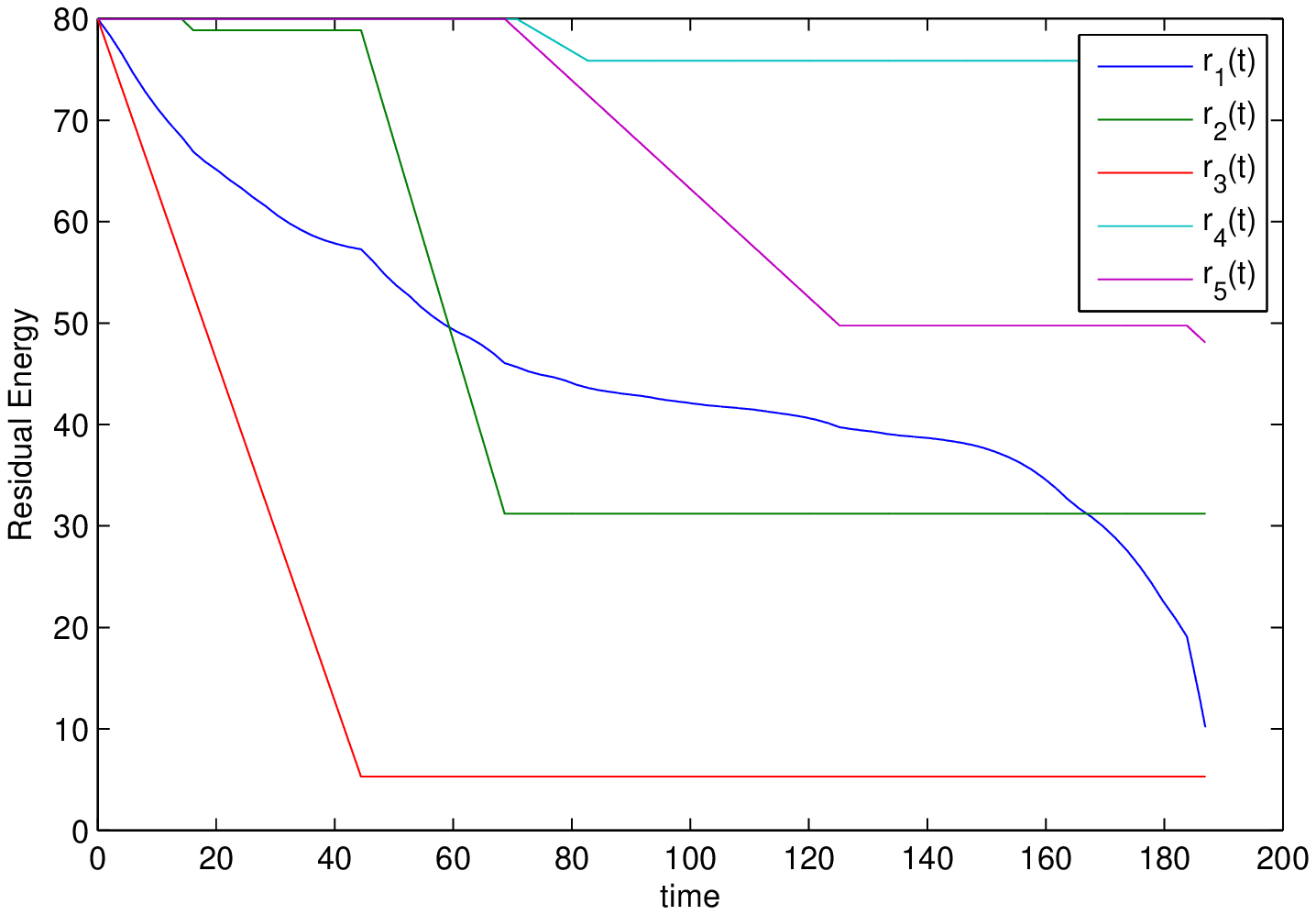}\caption{Residual energies over time
during the network lifetime}%
\label{res}%
\end{figure}

\begin{figure}[ptbh]
\centering
\includegraphics[scale=0.5]{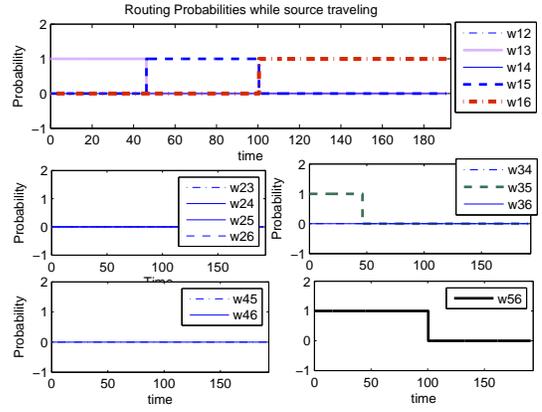}\caption{Optimal routing vectors for the
WSN in Fig. \ref{TPLG1} with a mobile source node (Problem III)}%
\label{RV_f3}%
\end{figure}

\begin{figure}[ptbh]
\centering
\includegraphics[scale=0.38]{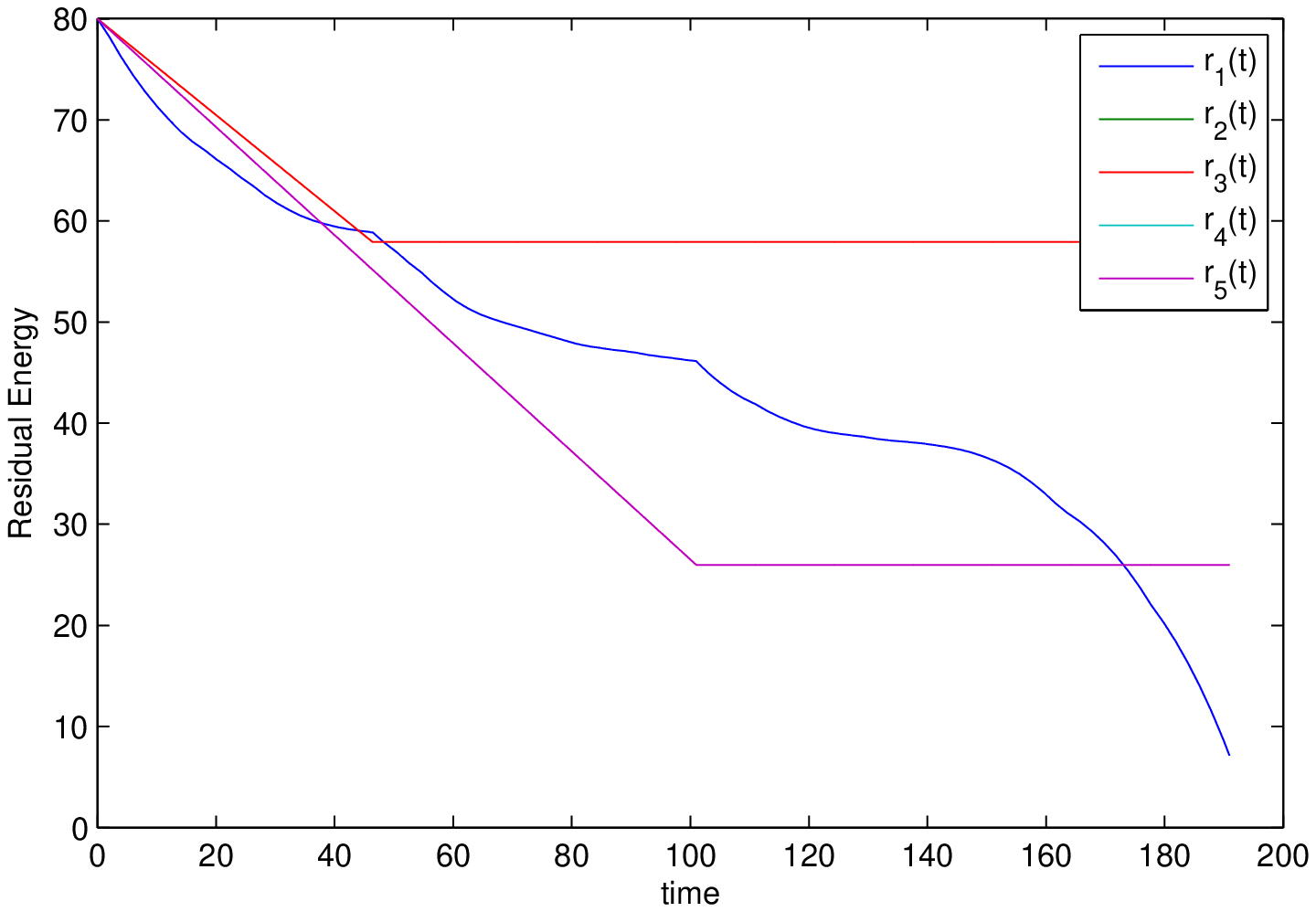}\caption{Residual energies over time
during the network lifetime}%
\label{resf3}%
\end{figure}

In this section, we use a WSN example to compare the performance of different
formulations based on the three different network lifetime definitions we have
considered. We consider a 6-node network as shown in Fig. \ref{TPLG1}. Nodes 1
and 6 are the source and base respectively, while the rest are relay nodes.
Let us set $C_{s}=0.0001,C_{f}=C_{r}=0.05$, and $\beta=2$ in the energy model.
We also set initial energies for the nodes $R_{i}=80$, $i=1,...,5$. Starting
with the source node at $(x_{0}(0),y_{0}(0))=(0,0)$, we solve the two
optimization problems (\ref{obj7})-(\ref{wij7}) with $\epsilon=1$ and the
equivalent shortest path problem of (\ref{obj8}) for Problems II and III
respectively as the trajectory of the source node evolves. Since  this trajectory is not known in advance,  in this
example we assume the source node moves based on a random walk as shown in Fig. 
\ref{TPLG1}. 
We first find the optimal routing vector by solving (\ref{obj7})-(\ref{wij7})
at each time step along the source node trajectory treating the network
topology as fixed for that step. Fig. \ref{RV} shows the routing vectors
during the network lifetime, i.e., the time when the source node depletes its
battery (in our numerical examples, we define this to be the time when the
source node energy level reaches $10\%$ of its initial energy).
The evolution of residual energies of all nodes during the network
lifetime is shown in Fig. \ref{res}. We can see that at $T=184$ the residual
energy of the source node drops to $10\%$ of its initial energy, hence that is
the optimal lifetime obtained using the definition where the soft constraint
$\sum_{i=1}^{N-1}r_{i}(T)$ is included in (\ref{obj4}). 
 Next, we use the WSN definition where $\sum_{i=1}^{N-1}r_{i}(T)=0$ is used as
a hard constraint. As already discussed, the corresponding problems
(\ref{obj9}) over the source node trajectory are generally infeasible.
Instead, we adopt the shortest path routing policy at each step to exploit
Theorem 3 with the understanding that the result (for this particular WSN
definition) is suboptimal. We consider the same source node trajectory as in
Fig. \ref{TPLG1}. The optimal routing vector updates are shown in Fig.
\ref{RV_f3}, while Fig. \ref{resf3} shows the residual energy of the nodes
during the network lifetime, which in this case is $T=188.49$, slightly longer
than the one obtained in Fig. \ref{res} with considerably less computational
effort. Also, note that since the source node always sends data packets
through the shortest path, it never uses nodes 2 and 4 for this particular
trajectory. As expected, (\ref{obj8})-(\ref{TerminalCondition2}) is not
feasible, however finding the shortest path at each step in fact improves the
network lifetime in the sense of the first time when the source node depletes
its energy. We point out, however, that this is not always the case and
several additional numerical examples show that this depends on the actual
trajectory relative to the relay node locations.

 Recall that $\epsilon$ is the weight of the soft constraint in problem $\mathbf{P2(}t\mathbf{)}$. Applying small or large  $\epsilon$ makes the problem closer to  $\mathbf{P1(}t\mathbf{)}$  or $\mathbf{P3(}t\mathbf{)}$ respectively, e.g. $\epsilon=0.5$ results in $T=190$,  however  $\epsilon=8$ shrinks the lifetime to $T=156$ suggesting that it is not optimal in this scenario to encourage the nodes to die simultaneously.
 
Based on the numerical results, it is obvious that the definition of a static WSN lifetime is not appropriate here. 
Finally, we observe that the routing vectors
are such that at each time step a subset of nodes is fully used ($w_{ij}=1$)
while the rest are not used at all. This suggests the possibility of a
\textquotedblleft bang-bang\textquotedblright\ type of optimal routing policy, which deserves further investigation.
\section{Optimal Control Formulation when source node trajectory is known in
advance}
\label{sec5} In this section, we consider the case when we have full advance
knowledge of the source node trajectory and include this information in the
optimal control problem. Defining the WSN lifetime to be the time when the source
node depletes its energy, i.e., using the definition in Problem I, Section
\ref{f1}, the problem is formulated as follows:%
\begin{gather}
\min_{w(t)}-\int_{0}^{T}dt\label{Obj_known}\\
\text{s.t. }\dot{r}_{i}(t)=-I_{i}(w(t)),\text{ }r_{i}(0)=R_{i},\ \ 
i=0,..,N-1\label{ri_known}\\%
\begin{bmatrix}
\dot{x}_{0}(t)\\
\dot{y}_{0}(t)\\
\end{bmatrix}
=%
\begin{bmatrix}
f_{x}(x_{0}(t),y_{0}(t))\\
f_{y}(x_{0}(t),y_{0}(t))\\
\end{bmatrix}, \ \ (x_0(0),y_0(0)) \text{ given}
\label{motion}\\
I_{i}(w(t))=G_{i}(w(t))[\sum_{j\in O(i)}w_{ij}(t)(C_{s}d_{i,j}^{2}%
+C_{f})+C_{r}],\nonumber\\
\quad\quad i=1,...,N-1\label{Ii_known}\\
I_{0}(w(t))=\sum_{j\in O(0)}w_{0j}(t)(C_{s}d_{0,j}^{2}(t)+C_{f}%
)\label{I0_known}\\
G_{i}(w(t))=%
{\displaystyle\sum\limits_{k\in I(i)}}
w_{ki}(t)G_{k}(w(t)),\text{ }i=1,..,N-1\label{Gi_known}\\
\sum_{j\in O_{i}}w_{ij}(t)=1,\quad0\leq w_{ij}(t)\leq1,i=0,\ldots
,N-1\label{w_known}\\
r_{0}(T)=0\label{R0T_known}\\
r_{0}(t)>0,\ t\in\lbrack0,T); \text{ }r_{i}(t)\geq0,\text{ }i=1,..,N-1,\text{ }t\in\lbrack0,T]
\label{riconstr_known}%
\end{gather}
where (\ref{motion}) specifies the trajectory of the source node. In this
problem, the state variables are the residual node energies, $r_{i}(t)$, as
well as the source node location at time $t$, $(x_{0}(t),y_{0}(t))$. Similar
to Section \ref{f1}, we obtain the Hamiltonian:
\begin{align}
&  H(w,t,\lambda)  =-1+\lambda_{0}(t)(-I_{0}(t))+\sum_{i=1}^{N-1}\lambda_{i}(t)(-I_{i}%
(t))+\nonumber\\
&  \lambda_{x}(t)f_{x}(x_{0}(t),y_{0}(t))+\lambda_{y}(t)f_{y}(x_{0}%
(t),y_{0}(t))\label{Ham_new}%
\end{align}
As before, $\lambda_{i}(t)$ is the costate corresponding to $r_{i}%
(t),\ i=0,...,N-1$ and we add $\lambda_{x}(t)$, $\lambda_{y}(t)$ to be the
costates of $x_{0}(t)$ and $y_{0}(t)$. Since we now know the equation of
motion for the source node in advance, this imposes terminal constraints for
the location of the source node at $t=T$. Thus, based on the dynamics in
(\ref{motion}) we can specify $x_{0}(T)$ and $y_{0}(T)$ as
$
x_{0}(T)=F_{x_{0}}(T)\text{ and }y_{0}(T)=F_{y_{0}}(T)$. 
Therefore, the terminal state constraint is:
\begin{align}
&  \Phi(\mathbf{r}(T),x_{0}(T),y_{0}(T))=\nonumber\\
&  \nu r_{0}(T)+\mu_{x}(x_{0}(T)-F_{x_{0}}(T))+\mu_{y}(y_{0}(T)-F_{y_{0}}(T))
\end{align}
where $\nu,\ \mu_{x},\ \text{and}\ \mu_{y}$ are unknown constants. It is
straightforward to show that $\lambda_{i}(t)$, $i=1,...,N-1$ are as in
(\ref{lambdaT1}). On the other hand, $\lambda_{x}$ and $\lambda_{y}$ must
satisfy:
\begin{align}
&  \dot{\lambda}_{x}(t)=-\dfrac{\partial H}{\partial x_{0}}=2C_{s}\lambda
_{0}(t)\sum_{j\in O(0)}[w_{0j}(t)(x_{0}(t)-x_{j})]\nonumber\\
&  -\lambda_{x}(t)\dfrac{\partial f_{x}}{\partial x_{0}}-\lambda_{y}%
(t)\dfrac{\partial f_{y}}{\partial x_{0}}\label{lambdadotN}\\
&  \dot{\lambda}_{y}(t)=-\dfrac{\partial H}{\partial y_{0}}=2C_{s}\lambda
_{0}(t)\sum_{j\in O(0)}[w_{0j}(t)(y_{0}(t)-y_{j})]\nonumber\\
&  -\lambda_{x}(t)\dfrac{\partial f_{x}}{\partial y_{0}}-\lambda_{y}%
(t)\dfrac{\partial f_{y}}{\partial y_{0}}\label{lambdadotN_1}%
\end{align}
with boundary conditions:
\begin{align}
&  \lambda_{x}(T)=\dfrac{\partial\Phi(\mathbf{r}(T),x_{0}(T),y_{0}%
(T))}{\partial x_{0}(T)}=\mu_{x}\label{lambdaNT}\\
&  \lambda_{y}(T)=\dfrac{\partial\Phi(\mathbf{r}(T),x_{0}(T),y_{0}%
(T))}{\partial y_{0}(T)}=\mu_{y}\label{lambdaN1T}%
\end{align}
The transversality condition $H(T)+\left.  \dfrac{\partial\Phi
}{\partial t}\right\vert _{t=T}=0$ gives:
\begin{align}
&  -1+\nu\dot{r}_{0}(T)+\lambda_{x}(T)\dot{x}_{0}(T)+\lambda_{y}(T)\dot{y}%
_{0}(T) +\nu\dot{r}_{0}(T)+\nonumber\\
& \mu_{x}\dot{x}_{0}(T)-\mu_{x}\dfrac{dF_{x_{0}}(T)}%
{dT}+\mu_{y}\dot{y}_{0}(T)
  -\mu_{y}\dfrac{dF_{y_{0}}(T)}{dT}=0\label{GenTrans}%
\end{align}
The solution of this problem is computationally challenging. Adjoining the control constraints (\ref{w_known}) to the Hamiltonian, the problem can be
numerically solved using a TPBVP solver, which, fortunately,  can be done off line in
advance of the source node initiating its known trajectory. It is also
possible that the solution is characterized by structural  properties (at least for some
$f_{x}(x_{0}(t),y_{0}(t))$, $f_{y}(x_{0}(t),y_{0}(t))$) which we plan to
investigate in followup work.

\section{Conclusions and future work}
We have redefined the lifetime for WSNs with a mobile source node to be the time until the source node runs out of energy. When the mobile node's trajectory is unknown in advance, we have shown that optimal routing vectors can be evaluated as solutions of a sequence of NLPs as the source node trajectory evolves. When the mobile node's trajectory is known in advance, we have limited ourselves to formulating an optimal control problem which requires an explicit off-line numerical solution. Ongoing work focuses on further exploring this case and on extensions to multiple mobile source nodes.  

\bibliographystyle{IEEEtran}
\bibliography{wxy,nx,renowang}

\end{document}